\documentclass[amssymb,aps,prl,twocolumn]{revtex4}
\usepackage{graphicx}

\begin{document}
\title{Photocurrents in nanotube junctions}
\author{D. A. Stewart}
\email[]{dstewar@sandia.gov}
\author{Fran\c{c}ois L\'{e}onard}
\affiliation{Sandia National Laboratories, Livermore, CA 94551}

\begin{abstract}
Photocurrents in nanotube {\it p-n} junctions are calculated using a
non-equilibrium Green function quantum transport formalism. The
short-circuit photocurrent displays band-to-band transitions and
photon-assisted tunneling, and has multiple sharp peaks in the infrared,
visible and ultraviolet ranges. The operation of such devices in the
nanoscale regime leads to unusual size effects, where the photocurrent
scales linearly and oscillates with device length. The oscillations can be
related to the density of states in the valence band, a factor that also
determines the relative magnitude of the photoresponse for different bands.
\end{abstract}

\pacs{85.60.-q, 72.40.+w, 73.63.Fg, 78.67.Ch}
\maketitle

Since their discovery\cite{Iijima}, Carbon Nanotubes (NTs) have been the
subject of intensive research due to their intriguing electronic and
structural properties, and have demonstrated great promise for future
nano-electronic devices\cite{nano}. However, their potential for
opto-electronic applications has received much less attention, despite the
ideal properties that NTs present. Indeed, a desirable property of
opto-electronic materials is a direct band-gap, since it allows optical
transitions to proceed without the intervention of phonons. NTs are unique
in this aspect, since {\it all} the bands in {\it all} semiconducting NTs
have a direct gap. Furthermore, the low dimensionality of NTs leads to a
diverging density of states at the band edge and a high surface-to-volume
ratio, reducing the sensitivity to temperature variations and allowing
efficient use of the material. Finally, non-radiative transitions can
significantly reduce the performance of conventional materials; NTs are
believed to have low defect density, reducing non-radiative transitions.

These unique properties of NTs are only starting to be explored for
opto-electronics. Recent experimental work\cite{misewich} has shown that NT
field-effect transistors can emit polarized light, while illumination of
these and other NT devices generates significant photocurrent\cite%
{freitag,narkis}. Such observations have also been made in semiconductor
nanowires\cite{wang,duan,gudiksen}. Often, the observed opto-electronic
effects are due to the presence of a {\it p-n }junction, with various
physical realizations (electrostatically defined\cite{misewich,freitag},
crossed-wire geometry\cite{duan}, or modulated chemical doping\cite{gudiksen}%
.) Modulated chemical doping of individual semiconducting single-wall NTs to
create ``on-tube'' {\it p-n} junctions has recently been reported in the
literature\cite{zhou}. These NT {\it p-n} junctions serve as excellent
testbeds for understanding opto-electronics at reduced dimensionality.

Here we show that in simple NT {\it p-n} junctions, the photocurrent shows
unusual features. Unlike traditional devices, the photoresponse in the NT
junctions consists of multiple sharp peaks, spanning the infrared, visible 
{\it and} ultraviolet ranges. Furthermore, at nanoscale dimensions, the NT
junctions show size effects, where the photocurrent scales and oscillates
with device length.

\begin{figure}
\begin{center}
\centering
\includegraphics[width=8.00cm]{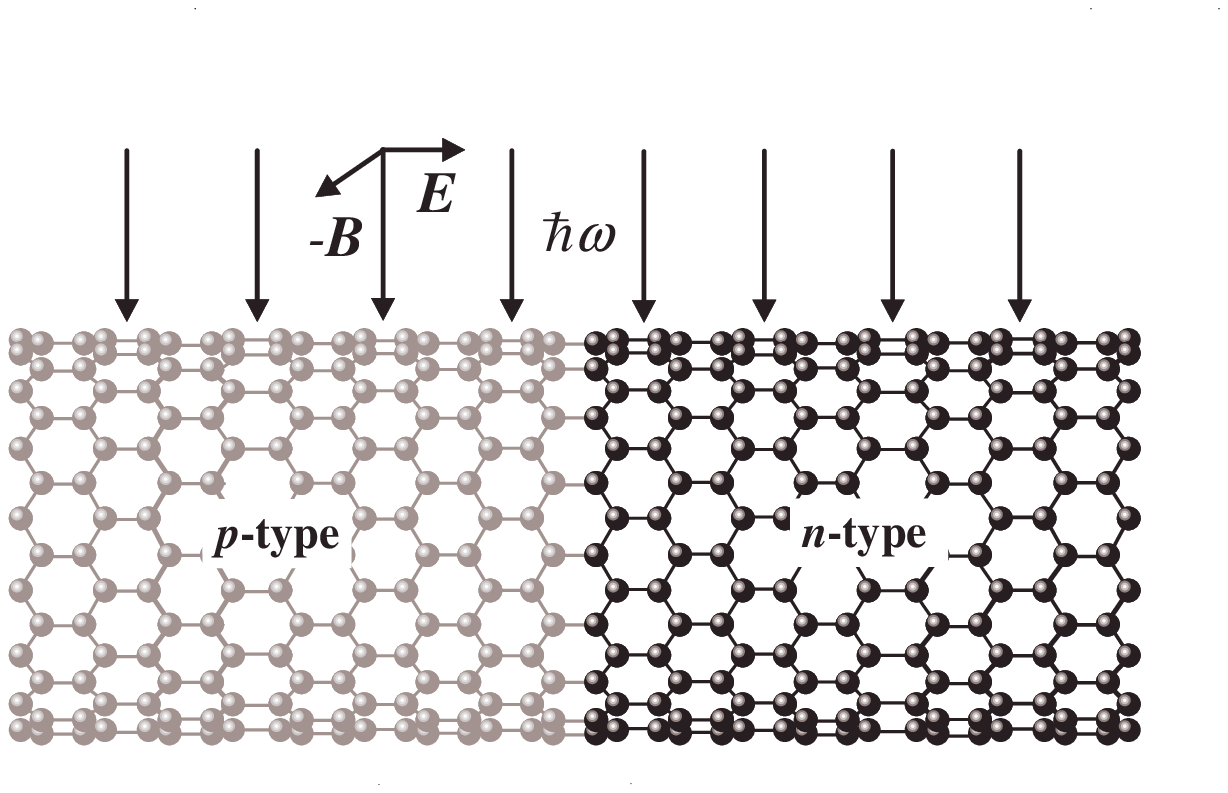}
\caption{Sketch of a portion of the scattering region for the (17,0) 
nanotube p-n junction.  The scattering region is attached to two 
semi-infinite leads (not shown in figure).  The incoming light of energy 
$\hbar\omega$ is polarized along the length of the nanotube, with electric 
field ${\bf E}$ and magnetic field ${\bf B}$.}
\label{nanotube_diagram}
\end{center}
\end{figure}

We consider a NT {\it p-n} junction under illumination, as shown in Fig. 
\ref{nanotube_diagram}.
We use a (17,0) single-wall, semiconducting NT of radius 0.66 nm, and model
the doping as in Ref. \cite{leonard}. Our tight-binding Hamiltonian with one 
$\pi $ orbital per carbon atom and nearest-neighbor matrix element of $%
\gamma =$ 2.5 eV gives a direct band-gap for this NT of 0.55 eV. The
incident light is assumed to be monochromatic, with polarization parallel to
the NT axis.

 To calculate the photocurrent, we use a non-equilibrium Green 
function formalism \cite{Datta}, in a real-space representation. The 
device consists of an illuminated scattering region connected to two dark 
semi-infinite leads, which are also doped (17,0) NTs. The structure of the 
(17,0) NT corresponds to parallel rings of 17 atoms, which are spaced by 
$0.07$ nm and 
$0.14$ nm. Each ring forms a layer in our system, with the rings in the
scattering region labelled from $1$ to $N$.

Since only layers $1$ and $N$ couple with the leads, the steady-state,
time-averaged current can be written as 
\begin{equation}
I=\frac{4e\gamma }{\pi \hbar }\int dE%
\mathop{\rm Re}%
\left[ G_{N+1,N}^{<}+G_{1,0}^{<}\right] .  \label{current}
\end{equation}%
(This expression assumes that the hole current is equal and opposite to the
electronic current and that the coupling of layers $1$ and $N$ to the leads
is $\gamma $. The appropriateness of this equation for time-dependent
Hamiltonians is discussed in Ref. \cite{anantram}) Here, $G_{N+1,N}^{<}$ and 
$G_{1,0}^{<}$ cross the scattering region boundary. To relate these terms to 
$G^{<}$ calculated entirely in the scattering region, we use the Dyson
equations\cite{lake} $\gamma G_{1,0}^{<}=G_{1,1}^{<}\Sigma
_{1,1}^{R0}+\Sigma _{1,1}^{<0}G_{1,1}^{R}$ and $\gamma
G_{N+1,N}^{<}=G_{N,N}^{<}\Sigma _{N,N}^{R0}+\Sigma _{N,N}^{<0}G_{N,N}^{R}$.
(We also note that in the presence of electron-photon interactions {\it and}
the potential step in the {\it p-n} junction\cite{anantram}, $%
G_{N+1,N}^{<}\neq G_{1,0}^{<}$, an equality that would normally hold for a
system without inelastic scattering.)

In the scattering region, $G^{<}$ is determined from the equations%
\begin{equation}
G^{<}=G^{R}\Sigma ^{<}G^{R\dagger }
\end{equation}%
and

\begin{equation}
G^{R}=\left[ EI-H-\Sigma ^{R}\right] ^{-1}\text{,}  \label{gr}
\end{equation}%
where $H$ is the Hamiltonian for the NT {\it p-n} junction under
illumination. The functions $\Sigma ^{R}$ and $\Sigma ^{<}$ represent the
interaction of the scattering region with the semi-infinite (17,0) NT leads
and the incident light. In the presence of the electron-photon interaction,
these functions depend on $G^{<}$, and the equations are coupled and
non-linear. To simplify the calculation, we perform an expansion to first
order in the light intensity: $G^{<}=G^{<0}+G^{<(ph)}$, $%
G^{R}=G^{R0}+G^{R(ph)}$, $\Sigma ^{<}=\Sigma ^{<0}+\Sigma ^{<(ph)}$ and $%
\Sigma ^{R}=\Sigma ^{R0}+\Sigma ^{R(ph)}$, where the order zero functions
represent the dark {\it p-n} junction. In the absence of an applied bias,
the only current in the device is the photocurrent 
\begin{equation}
I^{(ph)}=\frac{4e\gamma }{\pi \hbar }\int dE%
\mathop{\rm Re}%
\left[ G_{N+1,N}^{<(ph)}+G_{1,0}^{<(ph)}\right] ,
\end{equation}%
where%
\begin{eqnarray}
G^{<(ph)}&=&G^{R0}\Sigma ^{<(ph)}G^{R0\dagger }+G^{R0}\Sigma
^{R(ph)}G^{R0}\Sigma ^{<0}G^{R0\dagger } \nonumber \\
& &+G^{R0}\Sigma ^{<0}\left(
G^{R0}\Sigma ^{R(ph)}G^{R0}\right) ^{\dagger }.  \label{gaph}
\end{eqnarray}%
Here $G^{R0}=\left[ EI-H_{0}-\Sigma ^{R0}\right] ^{-1}$, with the
Hamiltonian $H_{0}$ and the self-energy function $\Sigma ^{R0}$ describing
the {\it p-n} junction without incident light, including the electrostatic
potential. (The Hamiltonian matrix elements are $%
H_{0}^{2l,2l-1}=H_{0}^{2l-1,2l}=2\gamma \cos \left( \frac{\pi J}{M}\right) $%
, $H_{0}^{2l,2l+1}=H_{0}^{2l+1,2l}=\gamma $, and $H_{0}^{l,l}=-eV_{l}$,
where $V_{l}$ is the electrostatic potential on layer $l$). The
self-energies $\Sigma ^{R0}$ due to the semi-infinite leads are calculated
using a standard iterative approach\cite{sancho}, and $\Sigma ^{<0}=-2f%
\mathop{\rm Im}%
\Sigma ^{R0}$ where $f$ is the Fermi function. The electrostatic potential
and charge distribution are determined using a self-consistent procedure as
in Ref. \cite{leonard}. The last two terms in Eq. $\left( \ref{gaph}\right) $
are proportional to $f(E)$, while, as discussed below, the first term is
proportional to $f(E-\hbar \omega )$. As long as the Fermi level is in the
bandgap and not too close to the conduction band edge, the ratio $f(E)/$ $%
f(E-\hbar \omega )$ is small, so we neglect the last two terms in Eq. $%
\left( \ref{gaph}\right) $. A benefit of this approach is that a model for $%
\Sigma ^{R(ph)}$ does not have to be formulated. (Including electron-hole
recombination and other multi-photon processes requires going beyond the
perturbation scheme introduced here, and is beyond the scope of this paper.
At low light intensities, our calculations should provide a reasonable
approximation.)

To derive the function $\Sigma ^{<(ph)}$ due to the electron-photon
interaction, we use the interaction Hamiltonian%
\begin{equation}
H_{el-ph}=\frac{e}{m}{\bf A}\cdot {\bf p}
\end{equation}%
where ${\bf A}${\bf \ }is the time-dependent electromagnetic vector
potential, ${\bf p}$ is the electronic momentum operator and $m$ is the
electron mass. Following the procedure of Ref. \cite{Henrickson}, we obtain%
\begin{equation}
\Sigma _{lm}^{<(ph)}(E)=\frac{2e^{2}a^{2}\gamma ^{2}F}{\hbar \omega
c\varepsilon }\sum_{pq}P_{lp}P_{qm}G_{pq}^{<0}(E-\hbar \omega ),
\label{sigma}
\end{equation}%
with%
\begin{equation}
P_{lm}=\left\{ 
\begin{array}{l}
-1\text{ \ }l-m=-1,\text{ }l\text{ even} \\ 
+1\text{ \ }l-m=1,\text{ }l\text{ odd} \\ 
-\cos \left( \frac{\pi J}{M}\right) \text{ \ }l-m=-1,\text{ }l\text{ odd} \\ 
+\cos \left( \frac{\pi J}{M}\right) \text{ \ }l-m=1,\text{ }l\text{ even} \\ 
0\text{ otherwise}%
\end{array}%
\right. .
\end{equation}%
In these equations, $a=0.07$ nm is the smallest ring separation, $\hbar
\omega $ is the photon energy, $c$ is the speed of light and $\varepsilon $
is the permittivity of free space. $J$ is the circumferential angular
momentum, labeling each of the $M=17$ bands of the NT. Conservation of $J$
is enforced, as required for light polarization parallel to the tube axis%
\cite{bozovic}. The photon flux $F$ represents the strength of the incoming
radiation, in units of photons per unit time per unit area. Since the
photocurrents calculated in this work are proportional to $F$, we focus on
the photoresponse $I/eF$. (This definition of the photoresponse might seem
awkward at first, having units of area/photon. However, the unusual size
effects in the nanotube devices are more transparent with this definition.)
Because $G^{<0}=-2f%
\mathop{\rm Im}%
G^{R0}$, $\Sigma _{lm}^{<(ph)}(E)$ is proportional to $f\left( E-\hbar
\omega \right) $.

\begin{figure}
\begin{center}
\centering
\includegraphics[width=8.0cm]{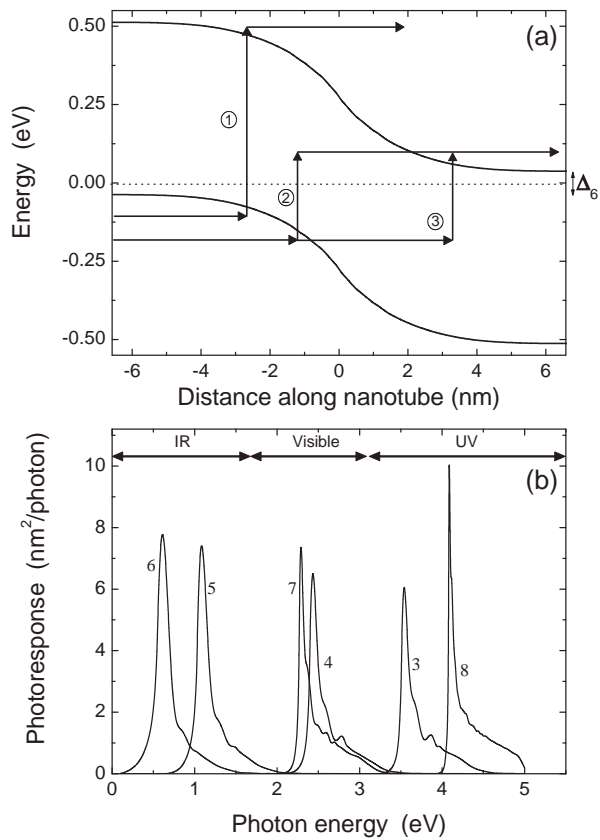}
\caption{Panel (a) shows the calculated self-consistent band diagram for 
the NT p-n junction.  Solid lines are the valence and conduction band 
edges for the $J = 6$ band.  Dotted line is the Fermi energ level.  Path 
$1$ represents a band-to-band transition while paths $2$ and $3$ 
represent photon-assisted tunneling.  Panel (b) shows the calculated 
photoresponse for this device, plotted seperately for each band $J$ 
(labelled with the value of $J$ in the figure.} 
\label{band_and_response} 
\end{center} 
\end{figure}

Figure \ref{band_and_response}a shows the calculated self-consistent band 
diagram for the NT {\it %
p-n} junction for a doping of $5\times 10^{-4}$ electrons/C atom. The figure
shows the conduction and valence band edges of the band with the smallest
band gap, $J=6$, shifted by the local electrostatic potential. The band
bending is characterized by a step at the junction and flat bands away from
the junction region\cite{leonard2}.

The calculated photoresponse for this device at zero bias when $128$ rings
are illuminated is shown in Fig.~ \ref{band_and_response}b. The 
photoresponse due to bands $%
J=6,5,7,4,3,8$ (increasing band gap) is separately plotted in the figure
(bands $11,12,10,13,14,9$ are equivalent). The general trend is for the
photoresponse for the different bands $J$ to peak at higher photon energies
as the band gap increases. Because the scattering cross-section decreases
with $\hbar \omega $ $\left[ \text{the }\left( \hbar \omega \right) ^{-1}%
\text{ factor in Eq.}~\left( \ref{sigma}\right) \right] $, one would expect
the maximum photoresponse attained for each band to decrease with band gap.
Surprisingly, the height of the peaks in Fig.~\ref{band_and_response}b 
does not follow this
behavior; in particular the response for band $J=8$ is actually larger than
the response for band $J=6$. This is a result of the different bands having
different effective masses. Indeed, the effective mass for $J=8$ is about 36
times larger than for $J=6$, leading to a much larger density of states near
the edge of the valence band. (The role of the density of states will be
discussed further below.)

Path 1 in Fig.~\ref{band_and_response}a shows a band-to-band transition 
with the absorption of a
photon of energy $\hbar \omega =0.6$ $eV$. An electron coming from the {\it p%
}-type side of the device in the valence band absorbs a photon and is
excited to the conduction band, where it then continues to the {\it n}-type
terminal. Such a transition is allowed when the photon energy exceeds the
band gap $E_{g}$ ($0.55$ $eV$ for band 6). Band-to-band photocurrents in the
NT device in the left lead due to electrons coming from the {\it n}-type
terminal vanish unless the photon energy is larger than the band gap plus
the potential step across the junction, $\hbar \omega \gtrsim 1$ $eV$ in
Fig.~\ref{band_and_response}b. This asymmetry in the currents to the left 
and right terminals
leads to the net photocurrents in Fig.~\ref{band_and_response}b.

While these band-to-band transitions explain part of the photoresponse, a
significant response exists at energies below the band gap. Such
contributions can be attributed to photon-assisted tunneling. Paths 2 and 3
in Fig.~\ref{band_and_response}a show two possible paths for 
photon-assisted tunneling. For a
given band $J$, this process can only occur when $\hbar \omega >\Delta _{J}$%
, where $\Delta _{J}$ is the difference between the asymptotic conduction
band edge on the {\it p}-type side and the asymptotic valence band edge on
the {\it n}-type side (equal to 0.06 $eV$ for the $J=6$ band shown in
Fig.~\ref{band_and_response}a). The photon-assisted tunneling thus turns 
on at $\hbar \omega
=\Delta _{J}$. As the photon energy increases above $\Delta _{J}$, more
states in the band gap become available for transport, and the photoresponse
increases. For the bands with larger band gaps, the tail due to
photon-assisted tunneling is less important relative to the band-to-band
peak, since tunneling probabilities decrease with band gap.

The photoresponse of the different bands leads to multiple sharp peaks in
three different regions of the electromagnetic spectrum: infrared, visible
and ultra-violet. This unusual behavior arises because all the bands in the
NT have a direct band-gap, which leads to a response over a wide energy
spectrum. The separation of this wide response into peaks grouped in
different regions of the electromagnetic spectrum is due to the particular
electronic band structure of the NT, which has groups of bands separated by
relatively large energies. The conduction band edges for $J=6$ and $J=5$
(infrared response) are separated from those of bands $4$ and 7 (visible
response) by about 0.6 eV, which are in turn separated by about 0.5 eV from
the $J=3$ and $J=8$ conduction band edges (ultraviolet response). Since the
photocurrent consists of an excitation from the valence band to the
conduction band, the separation in Fig.~\ref{band_and_response}b is twice 
these values.

\begin{figure}
\begin{center}
\centering
\includegraphics[width=8.0cm]{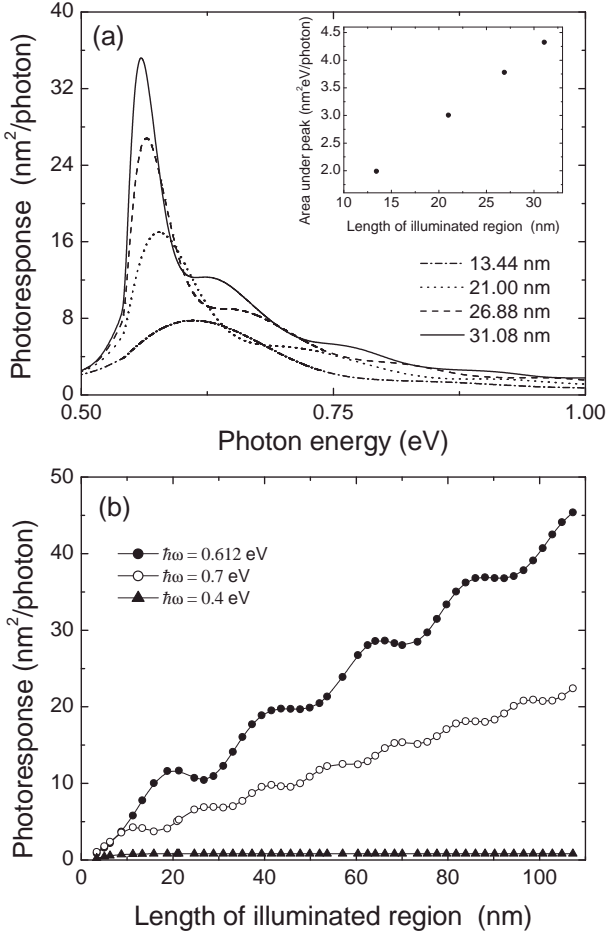}
\caption{(a) Dependence of the photoresponse for band $J=6$ on the length 
of the illuminated region.  The inset shows the integrated photoresponse, 
and its linear scaling with length.  The response for different photon 
energies is shown in (b).} 
\label{size_effects} 
\end{center}
\end{figure}

In conventional bulk junctions, the photoresponse depends only on the
dimensions of the device {\it perpendicular} to current flow. The nanotube
device however shows a dependence with {\it length}, as shown in 
Fig.~\ref{size_effects}.
Panel (a) in the figure shows the photoresponse of band $J=6$ for different
lengths of the illuminated region. As the illuminated region becomes longer,
the peak in the photoresponse increases and also moves to lower energies.
This behavior leads to linear scaling of the photoresponse with the length
of the illuminated region, as the integrated photoresponse in the inset of
Fig.~\ref{size_effects}a shows. The photoresponse for three different 
photon energies is
shown in Fig.~\ref{size_effects}b. Clearly for $\hbar \omega =0.4$ $eV$ 
(photon-assisted
tunneling) the photoresponse saturates with length, due to the fact that the
wavefunctions in the band gap decay exponentially away from the junction.
The response for $\hbar \omega =0.612$ $eV$ and $\hbar \omega =0.7$ $eV$
shows a completely different behavior, oscillating around a general linear
increase.

These surprising results can be understood from a simplified description of
the photocurrent in the nanotube. To this end, we note that the photocurrent
density flowing to the right lead $\left( I_{R}^{(ph)}=\frac{2e}{\pi \hbar }%
\int i_{R}(E)dE\right) $ can be expressed as%
\begin{equation}
i_{R}(E)=-\Gamma _{R}%
\mathop{\rm Im}%
\left( G^{R0}\Sigma ^{<(ph)}G^{R0\dagger }\right) _{NN}
\label{currentdensity}
\end{equation}%
where $\Gamma _{R}=-2i%
\mathop{\rm Im}%
\Sigma _{NN}^{R0}$. (Eq. $\left( \ref{currentdensity}\right) $ can be
obtained from Eq. $\left( \ref{current}\right) $ by using Dyson's equation $%
\gamma G_{N+1,N}^{<}=G_{N,N}^{<}\Sigma _{N,N}^{R0}+\Sigma
_{N,N}^{<0}G_{N,N}^{R}$ , the relation $G_{i,j}^{<}=-G_{j,i}^{<\ast }$ and
the condition that the net current must vanish for the dark junction at zero
bias.) At zero bias, $G^{<0}$ is purely imaginary, and therefore so is $%
\Sigma ^{<(ph)}$. The above equation becomes%
\begin{equation}
i_{R}(E)=-\Gamma _{R}\sum_{k,p}%
\mathop{\rm Im}%
\Sigma _{kp}^{<(ph)}\left\{ 
\mathop{\rm Re}%
G_{Nk}^{R0}%
\mathop{\rm Re}%
G_{Np}^{R0}+%
\mathop{\rm Im}%
G_{Nk}^{R0}%
\mathop{\rm Im}%
G_{Np}^{R0}\right\} .
\end{equation}%
For a given outgoing electron energy $E$ and photon energy $\hbar \omega $,
we have found numerically that the argument of the summation is peaked
around the diagonal $k=p$. Taking only these contributions, we obtain%
\begin{equation}
i_{R}(E)\sim -\Gamma _{R}\sum_{k}%
\mathop{\rm Im}%
\Sigma _{kk}^{<(ph)}\left| G_{Nk}^{R0}\right| ^{2}.  \label{ie}
\end{equation}%
The photocurrent in the NT can thus be understood in terms of the excitation
of electrons in each layer $k$ along the NT (the $%
\mathop{\rm Im}%
\Sigma _{kk}^{<(ph)}$ term in the above equation), and the subsequent
transmission to the lead by $G_{Nk}^{R0}$.

To explain the linear scaling, we note that for $\hbar \omega >E_{g}$ and $%
E>E_{c}^{-\infty }$ where $E_{c}^{-\infty }$ is the asymptotic value of the
conduction band edge on the {\it p}-type side of the device, there is a
section of the NT where band-to-band transitions are allowed, and the sum in
the last equation is dominated by this section of the NT. As the length of
the illumination region is increased, a longer section of the NT is
available for band-to-band transitions, leading to the linear scaling of the
current with length.

While this explains the linear scaling, there remains the questions of the
oscillations in the photoresponse and the dependence of the photoresponse on
the effective mass. To address this, we note that $%
\mathop{\rm Im}%
\Sigma _{kk}^{<(ph)}(E)$ contains terms like $%
\mathop{\rm Im}%
G_{kk}^{<0}(E-\hbar \omega )$, which are related to the density of states at
layer $k$, $D_{k}$, through $%
\mathop{\rm Im}%
G_{kk}^{<0}(E-\hbar \omega )=\pi f(E-\hbar \omega )D_{k}(E-\hbar \omega )$,
with $f(E)$ the Fermi function. Therefore, the photocurrent is sensitive to
the density of states at energy $E-\hbar \omega $. This explains the origin
of the dependence on the effective mass, and the relative height of the
peaks in Fig.~\ref{band_and_response}.

\begin{figure}
\begin{center}
\centering
\includegraphics[width=8.0cm]{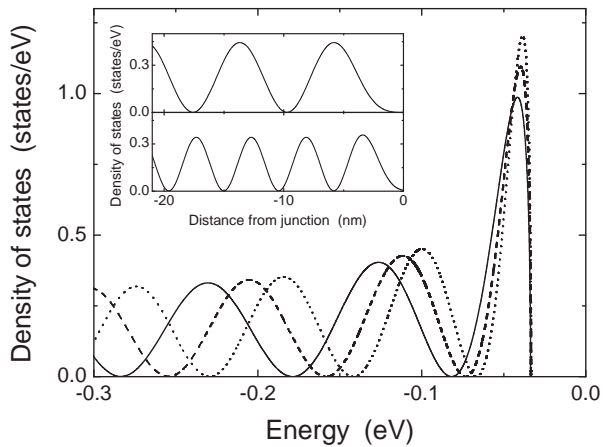}
\caption{Density of states at layer $11$ as a function of energy.  
Solid, dashed, and dotted lines are for illumination lengths of 24.78, 
26.88, 28.98 nm respectively.  The top (bottom) inset shows the density 
of states on the even rings at energy -0.1 eV (-0.2 eV).} 
\label{valence_dos} 
\end{center} 
\end{figure}

Figure \ref{valence_dos} shows the valence band density of states for 
$J=6,$ calculated at
layer $11$ for systems with illuminated lengths of $24.78$, $26.88$, and $%
28.98$ nm. (This is the density of states near the edge of the illuminated
region, which is moving further away from the junction as the length
increases). The density of states contains many peaks, and as the system
size changes, the peaks move in energy. Because the propagator $\left|
G_{Nk}^{R0}\right| ^{2}$ is sharply peaked at energy $E_{c}^{-\infty }$, the
photoresponse is particularly sensitive to the density of states at $%
E_{c}^{-\infty }-\hbar w$. At that energy, the density of states oscillates
as a function of the distance from the {\it p-n }junction, as illustrated in
the inset of Fig.~\ref{valence_dos}. This leads to the oscillations in 
the photoresponse as
a function of illumination length shown in Fig.~\ref{size_effects}b. The 
oscillation
wavelength of the density of states increases for energies closer to the
band edge, causing the different oscillation wavelengths for $\hbar w=0.612$ 
$eV$ and $\hbar w=0.7$ $eV$ in Fig.~\ref{size_effects}b.

Although the device presented here is fairly simple, it already shows the
richness of new phenomena that arises in nanoscale opto-electronics. One may
envision several uses for this or related devices (photodetection, power
generation, optical communication, etc.), but the important point is that
the behavior is much different from traditional devices and must be taken
into account in future device development.

This work was supported by the Office of Basic Energy Sciences, Division of
Materials Sciences, U.S. Department of Energy under contract No.
DE-AC04-94AL85000.


\begin{references}
\bibitem{Iijima} S. Iijima, Nature {\bf 354}, 56 (1991).

\bibitem{nano} A. Bachtold {\it et al.}, Science {\bf 294}, 1317 (2001); M.
S. Fuhrer {\it et al.}, Science {\bf 288}, 494 (2000); J. Kong {\it et al.},
Science {\bf 287}, 622 (2000).

\bibitem{misewich} J. A. Misewich {\it et al.}, Science {\bf 300}, 783
(2003).

\bibitem{freitag} M. Freitag {\it et al.}, Nano Lett. {\bf 3}, 1067 (2003).

\bibitem{narkis} T. R. Narkis {\it et al.}, in preparation.

\bibitem{wang} J. Wang {\it et al.}, Science {\bf 293}, 1455 (2001).

\bibitem{duan} X. Duan {\it et al.}, Nature {\bf 409}, 66 (2001).

\bibitem{gudiksen} M. S. Gudiksen {\it et al.}, Nature {\bf 409}, 66 (2002).

\bibitem{zhou} C. Zhou {\it et al.}, Science {\bf 290}, 1552 (2000).

\bibitem{leonard} F. L\'{e}onard and J. Tersoff, Phys. Rev. Lett. {\bf 85},
4767 (2000).

\bibitem{Datta} S. Datta, {\it Electronic transport in mesoscopic systems }%
(Cambridge University Press, Cambridge, 1995).

\bibitem{anantram} S. Datta and M.P. Anantram, Phys. Rev. B {\bf 45}, 13761
(1992).

\bibitem{lake} R. Lake {\it et al.}, J. Appl. Phys. {\bf 81}, 7845 (1997).

\bibitem{sancho} M.P. Lopez-Sancho, J.M. Lopez-Sancho, and J. Rubio, J.
Phys. F: Metal Phys. {\bf 15}, 851 (1985).

\bibitem{Henrickson} L. E. Henrickson, J. Appl. Phys. {\bf 91}, 6273 (2002).

\bibitem{bozovic} I. Bo\v{z}ovi\'{c}, N. Bo\v{z}ovi\'{c}, and M. Damnjanovi%
\'{c}, Phys. Rev. B {\bf 62}, 6971 (2000).

\bibitem{leonard2} The detailed properties of dark NT {\it p-n} junctions
are discussed in F. L\'{e}onard and J. Tersoff, Phys. Rev. Lett. {\bf 83},
5174 (1999). 

\end{references}
\end{document}